\documentstyle[prd,aps]{revtex}
\begin{document}
\input psfig
\pssilent
\title{Lorentz violations in canonical quantum gravity}
\author{Rodolfo Gambini}
\address{Instituto de F\'{\i}sica, Facultad de Ciencias,
Universidad 
de la Rep\'ublica, Igu\'a 4225, CP 11400 Montevideo, Uruguay}
\author{Jorge Pullin}
\address{Department of Physics and Astronomy, Louisiana State
University, 202 Nicholson Hall, Baton Rouge, LA 70803-4001}
\date{October 7th 2001}

\maketitle
\begin{abstract}
This is a summary of a talk given at the CP01 meeting on possible Lorentz 
anomalies in canonical quantum gravity. It briefly
reviews some initial explorations on the subject that have taken place
recently, and should be only be seen as a pointer to the literature
on the subject, mostly for outsiders.
\end{abstract}

\section{Introduction}

It has been recognized for some time that applying the rules of
quantum mechanics to general relativity is problematic
\cite{Rohi}. The fact that general relativity has a dimensionful
coupling constant immediately implies that a straightforward
perturbative approach will lead to a non-renormalizable theory. The
current majority-held point of view towards the problem of quantum
gravity is that general relativity is pathological and it has to be
replaced by a more general theory to be quantized. Current candidates
include ``M-Theory'', a non-perturbative superset of string theories
\cite{stri}.

A group of researchers has over the last decade devoted themselves to
reexamine the original question: is general relativity impossible to
quantize? The main rationale for this is that there is a feeling that
the question has not been examined in a detailed enough way. Yes,
apparent non-renormalizability suggests a problem, but it is not clear
if this is a problem of the theory itself, or of the way in which
perturbation theory is being implemented. A striking example of this
distinction is given by gravity in $2+1$ dimensions \cite{Ca}. In $2+1$
dimensions the Einstein equations imply space-time is locally
flat. The only possible degrees of freedom of gravity have to do with
global properties of the space-time. The resulting theory is not even
a field theory proper: its reduced phase space is finite
dimensional. Nevertheless, if one does a straight application of
perturbation theory to the quantization of general relativity in $2+1$
dimensions, the theory appears non-renormalizable for the same reasons
as gravity in $3+1$ dimensions. Witten was the first to point out that
a non-perturbative quantization should be possible, and after that was
accomplished, other researchers showed that one could indeed study the
theory perturbatively. The overriding question is: could we  be in
the same situation with gravity in $3+1$ dimensions? Other examples of
theories that do not appear to exist perturbatively but do
non-perturbatively are known \cite{DeWi}.

It is not surprising that quantizing general relativity should be
tricky. The theory is invariant under space-time diffeomorphisms,
reflecting the coordinate invariance of general relativity. Most
traditional quantum field theory techniques are not well suited for
diffeomorphism invariant theories, since they break such
invariance. We have only comparatively recently started to understand
how to quantize diffeomorphism invariant theories, but most of the
results have been for ``topological field theories'' (again, theories
with a finite number of degrees of freedom) \cite{top}. With almost no
exception, in general relativity is the first time we confront the
quantization of a theory that is both diffeomorphism invariant and
also has an infinite number of degrees of freedom.

Non-perturbative quantizations of general relativity (both using
canonical and path integral techniques) were first attempted in the
60's and 70's. Most of these attempts ended unsuccessfully.  The
expressions encountered either for the quantum Hamiltonian constraint
or the path integral were ill-defined and there was no idea of how to
regularize them. A revival of interest took place when Ashtekar
\cite{Asbook} discovered a new set of variables in terms of which
general relativity resembled a Yang--Mills theory. This opened the
possibility of applying techniques of non-perturbative quantization
that were successfully applied in the Yang--Mills context. The new
variables introduced by Ashtekar consist of replacing the spatial
metric (in canonical quantization, the theory is based on studying a
slice of space as it evolves in time) by a set of three vector fields
(triads), $E^a_i$ (the reader can pretend one is dealing with the
three electric fields of an $SU(2)$ Yang--Mills theory), and the
conjugate momentum behaves like an $SU(2)$ connection $A_a^i$. In
terms of these variables the constraints become polynomial and include
among them a Gauss law. That is, one can view the phase space of
general relativity as a submanifold of that of Yang--Mills theory.

Having the theory cast in terms of a Yang--Mills-like connection,
allows one to expand the wavefunctions in terms of traces of
holonomies (Wilson loops). The coefficients of the expansion
constitute the ``loop representation'', which had been studied for
Yang--Mills theories by Gambini and Trias \cite{GaTr} in the 80's and
was introduced into the gravitational context by Rovelli and Smolin
\cite{RoSm90}. Expressing things in terms of loops is quite attractive
in the gravitational context, since one can embody diffeomorphism
invariance in a remarkably clear way: one simply considers functions
of loops that are invariant under deformations of the loops. Such
functions have been studied by mathematicians since the time of Gauss,
and are called knot invariants \cite{RoSm88}.

Ashtekar and Lewandowski \cite{AsLe} introduced a set of mathematical
tools to deal with gauge-invariant diffeomorphism invariant functions
of a connection. These tools consist of a set of simple functions
(cylindrical functions) that can be rigorously endowed with a
measure. This is highly non-trivial, since it is a measure of
integration in an infinite-dimensional space. There are few examples
of such measures available. One now therefore has a definite,
mathematically controlled setting in which to construct a theory of
quantum gravity.

The basis of Wilson loops is overcomplete, and for many years this
caused problems when working in the loop representation. Rovelli and
Smolin \cite{RoSmspin} showed that one could find an elegant way of
labeling the independent elements of the basis of loops in terms of
``spin networks''. The latter are a mathematical construction first
introduced by Penrose consisting of embedded graphs of lines
intersecting at multivalent vertices, each line labeled by a
representation of $SU(2)$. The Ashtekar--Lewandowski measure is
particularly simple to understand in terms of spin networks.

Several attractive results have been obtained using these
techniques. For instance, the spectrum of the quantum operators
associated with the area of a surface and of the volume \cite{areavol}
of a region were studied in detail, and found to be
discrete. Expressions for the Hamiltonian constraint of gravity that
are finite and well defined were also proposed \cite{qsdvassil}. This
constitutes the first proposal ever of a non-trivial, well defined
theory of quantum gravity. The present efforts are concentrating on
demonstrating that the theory constructed has the correct classical
and semi-classical limit.

There has also been progress in applying these mathematical tools to
define a path-integral quantization of general relativity
\cite{spinfoam}. Impressive recent results have found a certain sense
in which the Lorentzian path integral can be made finite.

\section{Lorentz violations}

Why should there be Lorentz violations in non-perturbative quantum
gravity? The answer can only be tentative, since we do not have a
definite answer concerning the semi-classical limit of the
theory. Fortunately, progress in this area is happening at a
significant pace \cite{coherent} and I expect in near future CP
meetings a more definitive answer will be present.

To see why one might expect violations, let us consider at a very
broad and heuristic sense how would one construct a semi-classical
limit of the theory. When one has a quantum theory under control, one
should be able to build coherent states that are peaked around a
solution of the classical equations of motion, and to therefore study
quantum fluctuations. Since gravity is now under control, we can build
such states. However, there is an unusual twist when one carries out
this straightforward idea in the case of gravity. The twist is that
presumably one does not want to consider a coherent state that
approximates {\em any} classical metric. The reason for this is that
one can obviously construct many classical metrics that do not
correspond to a ``classical'' situation at all (for instance a metric
with gravitational waves of wavelength shorter than the Planck
length). Nothing prevents one from mathematically constructing such
coherent states. One should therefore introduce into the
semi-classical construction a proviso that one will only consider
states that approximate metrics that are ``reasonably smooth'' at a
certain length scale. This length scale is absolute. Therefore, unless
one elaborates further, Lorentz violations will occur.

As I stated before, this subject is rapidly evolving. Amelino-Camelia
has several articles on this point and I would refer the reader to
them \cite{Amelinolength}.

\section{Concrete calculations}

It is evidently that more work is needed before we can perform a
concrete calculation showing Lorentz violations. Up to present only a
handful of preliminary investigations have been carried out. In a
piece of work with Gambini \cite{GaPu} we studied the coupling of
gravity to Maxwell theory when gravity is treated quantum
mechanically. We took a na\"{\i}ve point of view in which we assumed
that the quantum state was given by a spin network state and studied
the coupling Hamiltonian pretending that the Maxwell fields were in a
semi-classical coherent state and only kept the leading terms. This
calculation is inadequate at many levels: first of all we did not
consider a quantum state that satisfies all constraints, and
considered a semi-classical limit of them. We just considered a
generic spin network state and treated the Maxwell theory as a regular
field theory. Under these assumptions, we showed that if the spin
network state had certain properties (namely it was parity-violating),
one got Lorentz violating corrections to Maxwell theory, that implied
a birefringence of space-time at the level of $\ell_p/\lambda$ where
$\ell_P$ is Planck's length and $\lambda$ is the wavelength of the
light. Such effect could influence the light arriving from gamma-ray
bursts and be within a few orders of magnitude of observation (very
much like the string inspired violations of Amelino-Camelia et
al. \cite{Amelinobi}). It was later noted that such birefringence is
severely constrained by optical observations of polarized sources
\cite{GlKo}. Alfaro et al. \cite{Alfarpho} have performed a
reexamination of these calculations, with a more careful motivation
for the quantum states considered, and found certain discrepancies in
the results with those of reference. The same authors \cite{Alfarofer}
have also extended these results to propagations of Fermions, finding
possibly observable effects in the time arrival of neutrinos. The
Fermionic calculations have a further drawback in that a given
expectation value for the connection has to be assumed in the quantum
state. The results are therefor highly dependent on this value, which
is not currently fixed by any a priori argument. There is rather
general consensus that effects at higher orders in $\ell_P/\lambda$
than the one we found are likely be present.  Such effects appear to
weak to be observed.

\section{Conclusions}

At present it appears possible that the canonical quantization of
gravity will, in its semi-classical limit, contain violations of
Lorentz symmetry. Some preliminary heuristic calculations have
exhibited this effect. The encouraging part is that rigorous
semi-classical studies are currently under way and it is expected that
in the next few years these questions will be given a definitive
answer.

\acknowledgements
This work was supported in part by grants NSF-PHY0090091, 
NSF-PHY-9800973, NSF-INT-9811610, and by funds of the Horace C. Hearne
Jr. Institute for Theoretical Physics.


\begin{references}
\bibitem{Rohi} For a historical review, see C. Rovelli gr-qc/0006061.

\bibitem{stri} There is a vast literature, 
including many good reviews on string theory. A standard reference is
the book by J. Polchinski, Cambridge University Press, Cambridge UK
(1998).


\bibitem{Ca} See for instance S. Carlip, 
{\em ``Gravity in $2+1$ dimensions''},
Cambridge University Press, Cambridge, UK (1998).

\bibitem{DeWi} B. DeWitt, Phys. Rev. {\bf 162}, 195 (1967).

\bibitem{top} See for instance D.~Birmingham, M.~Blau, 
M.~Rakowski and G.~Thompson,
Phys.\ Rept.\ {\bf 209}, 129 (1991).


\bibitem{Asbook} A. Ashtekar (Notes prepared in collaboration with R.
Tate), ``Lectures on non-perturbative canonical gravity'', Advanced
Series in Astrophysics and Cosmology Vol. 6, World Scientific,
Singapore (1991).

\bibitem{GaTr} R. Gambini, A. Trias, Nucl. Phys. 
{\bf B278}, 436 (1986).

\bibitem{RoSm90} C. Rovelli, L. Smolin, 
Nucl. Phys. {\bf B331}, 80 (1990).

\bibitem{RoSm88} C. Rovelli, L. Smolin, Phys. Rev. Lett. {\bf 61},
1155 (1988). 

\bibitem{AsLe} A. Ashtekar, J. Lewandowski, J. Math. Phys. {\bf 5},
2170 (1995).

\bibitem{RoSmspin} C. Rovelli, L. Smolin, 
Phys.\ Rev.\ {\bf D52}, 5743 (1995)

\bibitem{areavol} A.~Ashtekar and J.~Lewandowski,
Class.\ Quant.\ Grav.\  {\bf 14}, A55 (1997); Adv.\ Theor.\ 
Math.\ Phys.\  {\bf 1}, 388 (1998).


\bibitem{qsdvassil}C. Di Bartolo, R. Gambini, J. Griego, J. Pullin
Phys. Rev. Lett. {\bf 84}, 2314 (2000); Class. Quan. Grav. 
{\bf 17}, 3211-3237; 3239-3264 (2000). 

\bibitem{spinfoam} For a short review, see J. Baez,
Lect.Notes Phys.543:25-94, (2000).

\bibitem{coherent} T. Thiemann, Class. Quant. Grav. {\bf 18} 2025 (2001);
T. Thiemann, O. Winkler, Class. Quant. Grav. {\bf 18} 2561 (2001);
hep-th/0005234; hep-th0005235; H. Sahlmann, T. Thiemann, O. Winkler,
Nucl. Phys. {\bf B606}, 401 (2001); L. Bombelli gr-qc/0101080.

\bibitem{Amelinolength}
G. Amelino-Camelia, Phys. Lett {\bf B510}, 255 (2001); gr-qc/0106004.

\bibitem{GaPu} R. Gambini, J. Pullin, Phys. Rev. D 59, 124021 (1999).

\bibitem{Amelinobi} G. Amelino-Camelia, J. Ellis, N. Mavromatos,
D. Nanopoulos, S. Sarkar, Nature {\bf 393} 763 (1998). 

\bibitem{GlKo}R.~J.~Gleiser and C.~N.~Kozameh,
Phys.\ Rev.\ D {\bf 64}, 083007 (2001).

\bibitem{Alfarpho}
.~Alfaro, H.~A.~Morales-Tecotl and L.~F.~Urrutia,
hep-th/0108061.

\bibitem{Alfarofer}
J.~Alfaro, H.~A.~Morales-Tecotl and L.~F.~Urrutia,
Phys.\ Rev.\ Lett.\  {\bf 84}, 2318 (2000).

\end{references}
\end{document}